\documentclass[12pt]{article}
\usepackage[left=2cm,top=2cm,right=2cm,bottom=3cm]{geometry}
\usepackage{graphicx}
\usepackage{amsbsy,amssymb,amsmath,amscd,cite}

\usepackage[max6]{authblk}

\usepackage{caption,subcaption}

\newcommand{\Not}[1]{{\mathaccent'27 {#1}}}

\newcommand{\bc}{\begin{center}}
\newcommand{\ec}{\end{center}}

\newcommand{\ba}{\begin{array}}
\newcommand{\ea}{\end{array}}

\newcommand{\bt}{\begin{tabular}}
\newcommand{\et}{\end{tabular}}

\newcommand{\eems}{\end{multline*}}
\newcommand{\eem}{\end{multline}}
\newcommand{\eeams}{\end{eqnarray}\vspace{0.1cm}}
\newcommand{\eearms}{\end{eqnarray*}\vspace{0.1cm}}
\newcommand{\edmms}{\end{displaymath}\vspace{0.1cm}}

\newcommand{\bems}{\begin{multline*}}
\newcommand{\bem}{\begin{multline}}
\newcommand{\beams}{\vspace{0.1cm} \begin{eqnarray}}
\newcommand{\bearms}{\vspace{0.1cm} \begin{eqnarray*}}
\newcommand{\bdmms}{\vspace{0.1cm} \begin{displaymath}}

\newcommand{\be}{\begin{equation}}
\newcommand{\ee}{\end{equation}}

\newcommand{\bes}{\begin{equation*}}
\newcommand{\ees}{\end{equation*}}

\newcommand{\bea}{\begin{eqnarray}}
\newcommand{\eea}{\end{eqnarray}}

\newcommand{\bear}{\begin{eqnarray*}}
\newcommand{\eear}{\end{eqnarray*}}

\newcommand{\bdm}{\begin{displaymath}}
\newcommand{\edm}{\end{displaymath}}

\newcommand{\feop}{\hfill \rule{1.5mm}{1.5mm} \\}

\newcommand{\bit}{\begin{itemize}}
\newcommand{\eit}{\end{itemize}}

\newcommand{\ben}{\begin{enumerate}}
\newcommand{\een}{\end{enumerate}}

\newcommand{\bd}{\begin{document}}
\newcommand{\ed}{\end{document}}

\newcommand{\psp}{\hspace{0.6cm}}
\newcommand{\hspone}{\hspace{1cm}}
\newcommand{\hsptwo}{\hspace{2cm}}

\newcommand{\hspthree}{\hspace{3cm}}
\newcommand{\hspfour}{\hspace{4cm}}

\newcommand{\hspfive}{\hspace{5cm}}
\newcommand{\hspsix}{\hspace{6cm}}

\newcommand{\hspseven}{\hspace{7cm}}
\newcommand{\hspeight}{\hspace{8cm}}

\newcommand{\vspone}{\vspace{1cm}}
\newcommand{\vsptwo}{\vspace{2cm}}

\newcommand{\vspthree}{\vspace{3cm}}
\newcommand{\vspfour}{\vspace{4cm}}

\newcommand{\vspfive}{\vspace{5cm}}
\newcommand{\vspsix}{\vspace{6cm}}

\newcommand{\vspseven}{\vspace{7cm}}
\newcommand{\vspeight}{\vspace{8cm}}

\newcommand{\vspnine}{\vspace{9cm}}
\newcommand{\vspten}{\vspace{10cm}}

\newcommand{\vspeleven}{\vspace{11cm}}
\newcommand{\vsptwelve}{\vspace{12cm}}

\newcommand{\qett}{\mbox{$(q^{-1})$}}
\newcommand{\zett}{\mbox{$(z^{-1})$}}

\newcommand{\qo}{q^{-1}}
\newcommand{\zettu}{\mbox{$z^{-1}$}}

\newcommand{\linf}[1]{{#1}\rightarrow \infty}

\newcommand{\cf}[1]{(\ref{#1})}

\newcommand{\eiw} {\mathrm{e}^{\mathrm{i}\omega}}
\newcommand{\emiw}{\mathrm{e}^{-\mathrm{i}\omega}}

\newcommand{\tr}{\mbox{${\rm Tr}$}}
\newcommand{\var}{\mbox{${\rm Var}$}}
\newcommand{\diag}{\mbox{${\rm Diag}$}}
\newcommand{\cov}{\mbox{${\rm Cov}$}}
\newcommand{\Ex}{{\sf E}}

\newcommand{\erm}{\mathrm{e}}
\newcommand{\irm}{\mathrm{i}}
\newcommand{\Prj} {{\bf \Pi}^\bot}

\newcommand{\bA} {{\bf A}}
\newcommand{\bB} {{\bf B}}
\newcommand{\bC} {{\bf C}}
\newcommand{\bD} {{\bf D}}
\newcommand{\bE} {{\bf E}}
\newcommand{\bF} {{\bf F}}
\newcommand{\bG} {{\bf G}}
\newcommand{\bH} {{\bf H}}
\newcommand{\bI} {{\bf I}}
\newcommand{\bJ} {{\bf J}}
\newcommand{\bK} {{\bf K}}
\newcommand{\bL} {{\bf L}}
\newcommand{\bM} {{\bf M}}
\newcommand{\bN} {{\bf N}}
\newcommand{\bO} {{\bf O}}
\newcommand{\bP} {{\bf P}}
\newcommand{\bQ} {{\bf Q}}
\newcommand{\bR} {{\bf R}}
\newcommand{\bS} {{\bf S}}
\newcommand{\bT} {{\bf T}}
\newcommand{\bU} {{\bf U}}
\newcommand{\bV} {{\bf V}}
\newcommand{\bW} {{\bf W}}
\newcommand{\bY} {{\bf Y}}
\newcommand{\bX} {{\bf X}}
\newcommand{\bZ} {{\bf Z}}

\newcommand{\GLm} {{\bf \Lambda}}
\newcommand{\GSg} {{\bf \Sigma}}
\newcommand{\GPh} {{\bf \Phi}}
\newcommand{\GPs} {{\bf \Psi}}
\newcommand{\GPi} {{\bf \Pi}}
\newcommand{\GDl} {{\bf \Delta}}
\newcommand{\GGm} {{\bf \Gamma}}
\newcommand{\GUp} {{\bf \Upsilon}}
\newcommand{\GOm} {{\bf \Omega}}
\newcommand{\GTh} {{\bf \Theta}}
\newcommand{\GXi} {{\bf \Xi}}

\newcommand{\gal} {\boldsymbol{\alpha}}
\newcommand{\gbt} {\boldsymbol{\beta}}
\newcommand{\ggm} {\boldsymbol{\gamma}}
\newcommand{\gdl} {\boldsymbol{\delta}}
\newcommand{\gep} {\boldsymbol{\epsilon}}
\newcommand{\vep} {\boldsymbol{\varepsilon}}
\newcommand{\gzt} {\boldsymbol{\zeta}}
\newcommand{\get} {\boldsymbol{\eta}}
\newcommand{\gth} {\boldsymbol{\theta}}
\newcommand{\vth} {\boldsymbol{\vartheta}}
\newcommand{\git} {\boldsymbol{\iota}}
\newcommand{\gkp} {\boldsymbol{\kappa}}
\newcommand{\glm} {\boldsymbol{\lambda}}
\newcommand{\gmu} {\boldsymbol{\mu}}
\newcommand{\gnu} {\boldsymbol{\nu}}
\newcommand{\gxi} {\boldsymbol{\xi}}
\newcommand{\gpi} {\boldsymbol{\pi}}
\newcommand{\vpi} {\boldsymbol{\varpi}}
\newcommand{\grh} {\boldsymbol{\rho}}
\newcommand{\vrh} {\boldsymbol{\varrho}}
\newcommand{\gsg} {\boldsymbol{\sigma}}
\newcommand{\vsg} {\boldsymbol{\varsigma}}
\newcommand{\gtu} {\boldsymbol{\tau}}
\newcommand{\gup} {\boldsymbol{\upsilon}}
\newcommand{\gph} {\boldsymbol{\phi}}
\newcommand{\vph} {\boldsymbol{\varphi}}
\newcommand{\gch} {\boldsymbol{\chi}}
\newcommand{\gps} {\boldsymbol{\psi}}
\newcommand{\gom} {\boldsymbol{\omega}}

\newcommand{\Ba}  {{\bf a}}
\newcommand{\Bb}  {{\bf b}}
\newcommand{\Bc}  {{\bf c}}
\newcommand{\Bd}  {{\bf d}}
\newcommand{\Be}  {{\bf e}}
\newcommand{\Bf}  {{\bf f}}
\newcommand{\Bg}  {{\bf g}}
\newcommand{\Bh}  {{\bf h}}
\newcommand{\Bi}  {{\bf i}}
\newcommand{\Bj}  {{\bf j}}
\newcommand{\Bk}  {{\bf k}}
\newcommand{\Bl}  {{\bf l}}
\newcommand{\Bm}  {{\bf m}}
\newcommand{\Bn}  {{\bf n}}
\newcommand{\Bo}  {{\bf o}}
\newcommand{\Bp}  {{\bf p}}
\newcommand{\Bq}  {{\bf q}}
\newcommand{\Br}  {{\bf r}}
\newcommand{\Bs}  {{\bf s}}
\newcommand{\Bt}  {{\bf t}}
\newcommand{\Bu}  {{\bf u}}
\newcommand{\Bv}  {{\bf v}}
\newcommand{\Bw}  {{\bf w}}
\newcommand{\Bx}  {{\bf x}}
\newcommand{\By}  {{\bf y}}
\newcommand{\Bz}  {{\bf z}}

\newcommand{\za} {\boldsymbol{a}}
\newcommand{\zb} {\boldsymbol{b}}
\newcommand{\zc} {\boldsymbol{c}}
\newcommand{\zd} {\boldsymbol{d}}
\newcommand{\ze} {\boldsymbol{e}}
\newcommand{\zf} {\boldsymbol{f}}
\newcommand{\zg} {\boldsymbol{g}}
\newcommand{\zh} {\boldsymbol{h}}
\newcommand{\zi} {\boldsymbol{i}}
\newcommand{\zj} {\boldsymbol{j}}
\newcommand{\zk} {\boldsymbol{k}}
\newcommand{\zl} {\boldsymbol{\ell}}
\newcommand{\zm} {\boldsymbol{m}}
\newcommand{\zn} {\boldsymbol{n}}
\newcommand{\zo} {\boldsymbol{o}}
\newcommand{\zp} {\boldsymbol{p}}
\newcommand{\zq} {\boldsymbol{q}}
\newcommand{\zr} {\boldsymbol{r}}
\newcommand{\zs} {\boldsymbol{s}}
\newcommand{\zt} {\boldsymbol{t}}
\newcommand{\zu} {\boldsymbol{u}}
\newcommand{\zv} {\boldsymbol{v}}
\newcommand{\zw} {\boldsymbol{w}}
\newcommand{\zx} {\boldsymbol{x}}
\newcommand{\zy} {\boldsymbol{y}}
\newcommand{\zz} {\boldsymbol{z}}

\newcommand{\hide}[1] {}

\newtheorem{Thm}{Theorem}
\newtheorem{Pro}{Proposition}
\newtheorem{Defn}{Definition}
\newtheorem{Exm}{Example}
\newtheorem{Lem}{Lemma}
\newtheorem{Asm}{Assumption}
\newtheorem{Cor}{Corollary}
\newtheorem{Alg}{Algorithm}
\newtheorem{paRem}{Remark}
\newenvironment{Rem}{\begin{paRem}\rm }{\end{paRem}}

\newcommand{\ColvecTwo}[2]{\left[\ba{c} {#1} \\ {#2} \ea \right]}
\newcommand{\colvecTwo}[2]{\left[\ba{cc} {#1} & {#2} \ea \right]^\top}

\newcommand{\colvecThr}[2]{\left[\ba{ccc} {#1} & \cdots & {#2} \ea \right]^\top}
\newcommand{\ColvecThr}[2]{\left[\ba{c} {#1} \\ \vdots \\ {#2} \ea \right]}
\newcommand{\rowvecThr}[2]{\left[\ba{ccc} {#1} & \cdots & {#2} \ea \right]}
\newcommand{\MatThrTwo}[4]{\left[\ba{ccc} {#1}  & {#3} \\ \vdots & \vdots \\ {#2} & {#4} \ea \right]}

\usepackage{setspace}

\begin{document}

\title{Gaussian information matrix for Wiener model identification%
\thanks{
  The research is supported in part by the 
  Australian Research Council 
  under the contract DP130103909, by the Fund for Scientific 
  Research (FWO-Vlaanderen), by the Flemish Government (Methusalem), 
  the Belgian Government through the Inter university Poles of 
  Attraction (IAP VII) Program, and by the ERC advanced grant SNLSID, 
  under contract 320378.
}} 

\author[1]{Kaushik Mahata%
\thanks{Email: Kaushik.Mahata@newcastle.edu.au}}
\affil[1]{Department of Electrical Engineering, 
University of Newcastle, Callaghan, NSW-2308, Australia.}

\author[2]{Johan Schoukens%
\thanks{Email: Johan.Schoukens@vub.ac.be}}
\affil[2]{Department ELEC, Vrije Universiteit, Brussel, Building K, 
Pleinlaan 2, 1050, Brussels, Belgium.}

\date{\today}
\maketitle

\begin{abstract}
We present a closed form expression for the information matrix 
associated with the 
Wiener model identification problem under the assumption 
that the input signal is a stationary Gaussian process. This expression
holds under quite generic assumptions. We allow the linear sub-system 
to have a rational transfer function of arbitrary order, and the 
static nonlinearity to be a polynomial of arbitrary degree. We also 
present a simple expression for the determinant of the information 
matrix. The expressions presented herein has been used for optimal 
experiment design for Wiener model identification.
\\

\noindent
{\bf Keywords:} Wiener model identification, Information matrix, Gaussian 
input, Determinant.
\end{abstract}

\newpage

\section{Introduction}

In this paper we present an expression of the information matrix associated
with the Wiener model identification problem under a generic setting. 
Our analysis allows a rational model
the linear subsystem, and a polynomial nonlinearity. The
rational transfer function can be of arbitrary oder and a
nonlinearity with arbitrary polynomial order can be allowed. The
analysis assumes Gaussian stationary input process.

This work is mainly motivated by the experiment design problem. The
standard approach in experiment design is to choose the input 
excitation in order to optimize some monotonic function of the 
information matrix \cite{1924,IR-EE-RT_2007:003,
IR-EE-RT_2006:053,IR-EE-RT_2006:026,Bombois06,Hildebrand03}.
When the linear subsystem $G$ of the underlying Wiener 
system has a generic rational transfer function, it
has an infinite impulse response. Consequently, the information matrix
$\bJ$ becomes a function of the of higher order joint moments of the 
input process $u(t),u(t-1),u(t-2), \ldots$. It is very challenging to 
optimize any criterion of $\bJ$ with respect to the all higher order 
moments of potentially infinitely many consecutive samples of the input
$u(t)$. In fact, it is   
quite difficult to just compute $\bJ$. Firstly, the available formulae 
for calculating higher order moments are quite challenging to program. 
More importantly, the complexity of the resulting algorithm 
typically grows exponentially with the length of the impulse response of 
$G$ \cite{Kan_08}. In fact, to the best of our knowledge
no previous authors have considered handling this issue when $G$ is not a
finite impulse response system. 
Even when a finite impulse response system is
considered in the literature, the order of the system have 
been restricted to 4 or less. In fact, when compared with the 
experiment design literature for linear system identification 
\cite{1924,IR-EE-RT_2007:003,
IR-EE-RT_2006:053,IR-EE-RT_2006:026,Bombois06,Hildebrand03},
the the line of research in the nonlinear input design has undergone a 
significant paradigm shift.
Most of the  preliminary studies reported so far 
\cite{Forgione_14,Gevers_12,Hjalmersson_07,Larsson_10,Valenzuela_13},
have considered a deterministic setting.
Among these
the multi-level excitation approach \cite{Forgione_14,DeCock_13,DeCock_14,
Larsson_10} appears to be popular lately. These deterministic 
methods do have their limitations.  
The multi-level approach is often not tractable when we
increase the number of levels. The majority of these methods are unable to 
handle IIR-type non-linear systems.

We show when the input process is Gaussian there is a 
simple algorithm to compute $\bJ$. This analysis also reveals some 
interesting mathematical structures, that allows us to parameterize the 
set of all admissible information matrices with a finite number of
parameters. Hence the experiment designer needs to solve only a 
finite dimensional problem. See \cite{Mahata15} for the details of 
how the expressions presented herein can be used in experiment design.


\section{Information matrix and its determinant}

In this section we present our main findings about the information 
matrix $\bJ$ and its determinant. We start 
in Section \ref{PRELIM} with the basic notation and introduce the 
formal problem setting. In particular, we introduce a generalized 
framework for setting up the constraint to ensure unique 
identifiability of the 
Wiener model. Next in Section \ref{RATIONAL} we list the main 
results. In particular we use a state space representations of 
underlying transfer functions. We believe this approach simplifies
the analysis, and illuminates the underlying mathematical structures 
to a significant extent.

\subsection{Model parameterization and identifiability}
\label{PRELIM}

A Wiener system is a cascade of a linear
time invariant system followed by a static nonlinearity. We 
assume that the linear sub-system has a rational transfer function 
\be
G(z,\gth) = 
\frac{g_0 + g_1 z^{-1} + \cdots + g_n z^{-n} }
{1 + a_1 z^{-1} + \cdots + a_n z^{-n} },
\label{Gparameterization}
\ee
parameterized by the parameter vector $\gth$ defined as
\be
\gth = [ \ 
a_1 \ \ \cdots \ \ a_n  \ \
g_1 \ \ \cdots \ \ g_n  \ \ g_0 \ ]^{\intercal}. 
\ee
The output of the linear model is denoted by $w$:
\be
w(t,\gth) = G(z,\gth) u(t).
\label{EQN1}
\ee
The static nonlinearity is modeled by a polynomial 
$\wp$ of order $m$:
\[
\wp(x,\bar{\gal}) = 
\alpha_0 + \alpha_1 x + \cdots + \alpha_m x^m,
\] 
parameterized by the vector of polynomial coefficients 
\[
\bar{\gal}
= [ \ \alpha_0  \ \ \alpha_1 \ \ \cdots \ \ \alpha_m \ 
]^{\intercal}.
\]
Therefore, the Wiener model equation takes the form
\be
M(\vth,\Bu_t) = \wp \{ G(z,\gth) u(t), \bar{\gal} \}. 
\label{MODEL}
\ee
It is tempting to choose $\vth = [ \ \bar{\gal}^{\intercal}
\ \ \gth^{\intercal} \ ]^{\intercal}$. But this parameterization
fails to ensure unique identifiability.  
We cannot allow all the 
components of $\gth$ and $\bar{\gal}$ to vary freely
while remaining independent of each other. 
The reason is straightforward. The transfer operator
between $u$ and $y$ does not change by  dividing $G$ by a scalar 
$\varrho \ne 0$, and  multiplying $\alpha_k$ 
by $\varrho^k$ for all $k=1,2,\ldots,m$. 
For this reason we must impose
some additional constraint on the parameters. 
In this paper we allow
varying the static gain of $G$ freely, and impose a normalization
constraint on $\bar{\gal}$.

\begin{Asm}
\label{ASM2}
There is a known vector
\be
 \gup = [ \ \upsilon_0 \ \ \upsilon_1 
 \ \ \cdots \ \ \upsilon_m \ ]^{\intercal}
\label{DEFINITION_GUP}
\ee
such that
\be
\alpha_0 \upsilon_0 + 
\alpha_1 \upsilon_1 +
\cdots + 
\alpha_m \upsilon_m
=
1,
\label{CONSTRAINT}
\ee
where $\upsilon_{\ell} \ne 0$ for some known
$\ell \in \{ 1,2,\ldots,m \}$. 
\end{Asm}

The choice of $\ell$ is often governed by the prior knowledge on the
type of nonlinearity. Typically $\ell \ne 0$, because it is often 
the case that $\alpha_0 = 0$. For an odd (even) nonlinearity $\ell$
must be an odd (even) number. In our experience, the choice of $\ell$
does not influence the asymptotic large sample 
accuracy of the estimated model.
%

\begin{Exm} It is common to take $\gup = (0,1,\ldots,0)$ or
$\gup = (0,\ldots,0,1)$. Another possibility would be to take 
$\gup = (1,\ldots,1)$ implying $\wp(1) = 1$. Note that the 
choice $\gup = (1,0,\cdots,0)$ is forbidden. It leads to a model
that is not identifiable.
\end{Exm}

Since $\upsilon_{\ell} \ne 0$ under Assumption \ref{ASM2}, 
we can rewrite \cf{CONSTRAINT} as
\be
\alpha_{\ell} = 
\frac{ 1 }{ \upsilon_{\ell} }
\left \{
1
-
\sum_{ \scriptsize{ \ba{c} k=0 \\ k \ne \ell \ea} }^{m} 
\upsilon_k \alpha_k
\right \}. 
\label{ALPHA_L}
\ee
Equation \cf{ALPHA_L} can be built into the identification algorithm.
We do not identify $\alpha_{\ell}$ separately, but express
it using \cf{ALPHA_L}. We define the parameter vector 
\be
\gal := [ \ \alpha_{i_1} 
\ \ \cdots \ \ \alpha_{i_{m}} \ ]^{\intercal},
\label{STRANGE_INDICES}
\ee
where the indices $i_k \in \{ 0,1,\ldots,m \}$ are chosen such that 
$i_k \ne \ell$ for all $k$, and $i_k \ne i_j$  whenever $k \ne j$.
Note that mapping $k \rightarrow i_k$ is quite flexible, and we need 
not impose any further restriction on this mapping.
The identification algorithm estimates 
\[
\vth = [ \ \gal^{\intercal} \ \  \gth^{\intercal} \ ]^{\intercal}
\]
from the data. Defining 
\begin{align}
\nonumber
\bL 
&= 
\left[ \ba{ccccc} 
1 & 0 & \cdots & 0 & - \upsilon_{i_1}/\upsilon_{\ell} \\
0 & 1 & \cdots & 0 & - \upsilon_{i_2}/\upsilon_{\ell} \\
\vdots & \vdots & & \vdots & \vdots \\
0 & 0 & \cdots & 1 & - \upsilon_{i_{m}}/\upsilon_{\ell} \ea \right], 
\\
\bP 
&= 
[ \ \ze_{i_1} \ \ \cdots \ \ \ze_{i_m} \ \ \ze_{\ell} \ ]^{\intercal},
\label{DEFINITION_PERMUTATION}
\end{align}
with $\ze_k$ denoting the $k+1$ th column of $(m+1) \times (m+1)$ 
identity matrix, it can be verified from  \cf{ALPHA_L} that 
\be
[ \ \gal^{\intercal} \ \ \alpha_{\ell} \ ]^{\intercal}
=
\bP \bar{\gal}
=
\bL^{\intercal} \gal.
\label{EXPRESSOIN_GAL}
\ee


\subsection{Main theoretical results}
\label{RATIONAL}



Let $\Ba = [ \ a_1 \ \ \cdots \ \ a_n \ ]^{\intercal}$, and
$\Bg = [ \ g_1 \ \ \cdots \ \ g_n  \ ]^{\intercal}$. Then
we can write \cf{Gparameterization} as 
\be
G(z,\gth) 
=
g_0 + (\Bg - \Ba g_0)^{\intercal}(z \bI - \bA_1)^{-1} \Bb_1,
\label{A5a}
\ee
where $(\bA_1,\Bb_1)$ is in controllable canonical form, {\em i.e.} 
\be
\bA_1 =
\left[ \ba{cccc}
-a_1 & \cdots & -a_{n-1} & -a_{n} \\
1 & \cdots & 0 & 0 \\
\vdots & \ddots & \vdots & \vdots \\
0 & \cdots & 1 & 0 \ea \right],
\
\Bb_1 = \left[ \ba{c} 1 \\ 0 \\ \vdots \\ 0 \ea \right].
\label{A.5}
\ee
Note that we can impose the structure \cf{A5a} and \cf{A.5} without 
any loss of generality. We make the following assumption throughout 
the paper, where $\Not{\gth}$ denotes the true value of $\gth$.

\begin{Asm} $G(z,\Not{\gth})$ is asymptotically stable. Consequently, all
the eigenvalues of $\Not{\bA}_1$ (which denotes the true value of
$\bA_1$) are located inside the unit disc in the
complex plane. In addition, the state space realization \cf{A5a} is 
minimal.
\label{asm_stable}
\end{Asm}

\begin{Lem} Define the matrices $\bA, \Bb, \Bc$ and $\bar{\bC}$ as
\begin{align}
\nonumber
\bar{\bC} &=
\left[ \ba{ccc}
\bI & -\Not{g}_0 \bI & 0 \\
0 & \bI & 0 \\
0 & -\Not{\Ba}^{\intercal} & 1 
\ea \right],
\  
\Bb = 
\left[ \ba{c} {\bf 0}_{n \times 1} \\ {\bf 0}_{n \times 1} \\ 1 \ea \right],
\Bc = 
\left[ \ba{c} {\bf 0}_{n \times 1} \\ \Not{\Bg} \\
\Not{g}_0
\ea \right],
\\
\bA &= 
\bar{\bC}
\left[ \ba{ccc} 
\Not{\bA}_1 & -\Bb_1 (\Not{\Bg} - \Not{\Ba} \Not{g}_0)^{\intercal} 
& {\bf 0}_{ n \times 1} \\
{\bf 0}_{n\times n} & \Not{\bA}_1 & \Bb_1 \\ 
{\bf 0}_{ 1 \times n} & {\bf 0}_{ 1 \times n} & 0 
\ea \right]
\bar{\bC}^{-1},
\label{A2b2C2}
\end{align}
where $\Not{\bA}_1, \Not{g}_0$, etc  are obtained by setting 
$\gth = \Not{\gth}$ in $\bA_1, g_0$, etc. Consider the stochastic process
$\Bx$ which is given in state space form as
\begin{align}
\Bx(t) &= \bA \Bx(t-1) + \Bb u(t). 
\label{STATESPACE}
\end{align} 
Then $w(t,\Not{\gth}) = \Bc^{\intercal} \Bx(t)$, and 
\be
\Bv_t 
= 
\left[ \ba{c} 
\bL \bP \Bz(t,\Not{\gth}) \\ 
\Bx(t) \gal_2^{\intercal} \Bz(t,\Not{\gth}) 
\ea \right],
\label{Expression_vt}
\ee
where we define 
\be
\Bz(t,\gth)
:=
[ \
1 \ \
w(t,\gth)
\ \
\{ w(t,\gth) \}^2
\ \
\cdots
\ \
\{ w(t,\gth) \}^m \ ]^{\intercal},
\label{ExpressionJ}
\ee
\be
\gal_2 = [ \ \Not{\alpha}_1 \ \ 2 \Not{\alpha}_2 \ \ 
\cdots m \Not{\alpha}_m \ \ 0 \ ]^{\intercal},
\label{DEFINITION_GAL_2}
\ee
with $\Not{\alpha}_k$ being the true value of $\alpha_k$.
\label{LEM2}
\end{Lem}

\noindent
{\bf Proof:}  See Section \ref{PROOF_OF_LEM2}.
\feop

\begin{Rem} Lemma \ref{LEM2}  does  
cover the case when $G$ is of finite impulse response
type, i.e.,
\[
G(z,\gth) = g_0 + g_1 z^{-1} + \cdots + g_n z^{-n}.
\]
In this case $\gth = [ \ \Bg^{\intercal} \ \ g_0 ]^{\intercal}$,
and $\Ba = 0$. The expressions \cf{A5a} and \cf{A.5} still hold with 
$\Ba = 0$. While finding a realization of $G_1$ we do not need 
to consider the derivatives with respect to $\Ba$. As a result we
get 
\[
\bA = \left[ \ba{cc} \Not{\bA}_1 & \Bb_1 \\ 
{\bf 0}_{1 \times n} & 0 \ea \right],
\ \ \
\Bb = \left[ \ba{c} {\bf 0}_{n \times 1} \\ 1 \ea \right],
\]
$\bar{\bC} = \bI$ and $\Bc = \gth$.
\end{Rem}

The consequence of  Lemma \ref{LEM2} is that 
$\bJ = \Ex \{ \Bv_t \Bv_t^{\intercal} \}$ is a function of the 
moments of the state vector $\Bx$. For the purpose of setting up an 
input design problem we can parameterize $\bJ$ in terms of the
moments of the random vector $\Bx$. In particular, when $u(t)$ is
Gaussian, then so is $\Bx(t)$. Hence for a Gaussian input $\bJ$ 
is a function of the mean and the covariance matrix of $\Bx(t)$.
As the next Theorem reveals, we can obtain a closed form expression
for $\bJ$.

\begin{Asm}
The input process $u(t)$ is stationary Gaussian with mean 
$\eta_u$. 
\label{ASM3}
\end{Asm}


Under Assumption \ref{ASM3}, $\Bx$ is a Gaussian random vector with 
mean 
\be
\get := \Ex \{ \Bx (t) \} =  (\bI - \bA)^{-1} \Bb \eta_u.
\ee
Let us define
\be
\GSg = \Ex \{ [ \Bx(t) - \get ] [ \Bx(t) - \get ]^{\intercal} \}.
\label{DEFINITION_SIGMA}
\ee
Consequently, $\Bc^{\intercal} \Bx(t)$ is a Gaussian random variable 
such that
\begin{subequations}
\begin{align}
\gamma &:= \Ex \{ \Bc^{\intercal} \Bx(t) \} = \Bc^{\intercal} \get.
\\
\sigma &:=  \Ex \{ \Bc^{\intercal} \Bx(t) - \gamma \}^2
= \Bc^{\intercal} \GSg \Bc.
\end{align}
\label{gamma_sigma}
\end{subequations}
In the rest of the paper we denote
\[
\GLm := 
\Ex \{ \Bz(t,\Not{\gth}) [\Bz(t,\Not{\gth})]^{\intercal} \}.
\]

\begin{Rem}
It is possible to express $\GSg$ as well in terms of $\bA$, $\Bb$,
and the  power spectral density $\Phi$ of $u$. However, we postpone 
that for a while. We first express $\bJ$ in terms of $\GSg$ and 
$\get$, and later connect $\Phi$ with $\GSg$.
This approach  suits the purpose of input 
design, where it is simpler to work with a parameterization of $\GSg$ 
than to work with $\Phi$ directly.
\end{Rem}

\begin{Rem}
The correlation matrix 
$\GLm $
can be given entirely as a function of
the mean $\gamma$ and variance $\sigma$
of $\Bc^{\intercal} \Bx(t)$. 
Many different ways are used in the literature to 
express the moments of the scalar valued normal density. 
There are some explicit
formulae for smaller orders. 
We find it convenient to use a recursive
formula in the implementation. 
Let us denote $\mu_k(\gamma,\sigma) :=
\Ex \{ (\Bc^{\intercal} \Bx )^k \}$. 
So $\mu_k$ is a 
function of $\sigma$ and $\mu$. Then 
$\mu_k(\gamma,\sigma)$ satisfies the recursion
\cite[Chapter 5]{Papoulis_91}:
\be
\mu_k( \gamma, \sigma) = 
\gamma^k +
\frac{k(k-1)}{2} \int_{0}^{\sigma}
\mu_{k-2}( \tau, \sigma) \
\mathrm{d} \tau.
\label{RECURSSION}
\ee
Note that the recursion \cf{RECURSSION} 
needs to be carried out separately for 
even and odd values of $k$. For even 
valued $k$ one can initialize the recursion
with $\mu_0(\gamma,\sigma) = 1$, and for 
the odd values of $k$ we initialize with
$\mu_1(\gamma,\sigma) = \gamma$. This allows us to form 
\begin{align}
\nonumber
\GLm &= 
\left[ \ba{cccc}
\mu_0(\gamma,\sigma)   & \mu_1(\gamma,\sigma)  & \cdots & 
\mu_m(\gamma,\sigma)  \\
\mu_1(\gamma,\sigma)   & \mu_2(\gamma,\sigma)  & \cdots &  
\mu_{m+1}(\gamma,\sigma)  \\
\vdots & \vdots &  & \vdots \\
\mu_m(\gamma,\sigma)   & \mu_{m+1}(\gamma,\sigma)   & \cdots &  
\mu_{2m+1}(\gamma,\sigma) 
\ea
\right].
\label{A26a}
\end{align}
\end{Rem}

Since $\Bx$ is Gaussian, all the moments of 
$\Bx$ can be expressed as  functions of $\get$ and $\GSg$. 
This allows us to derive manageable expressions of $\bJ$ as a
function of  $\get$ and $\GSg$. This is shown next. 

\begin{Thm}
Define 
\be
\gal_1 = [ \ 0 \ \ \Not{\alpha}_1, 
 \ \ 
2 \Not{\alpha}_2 \ \ \cdots \ \ m \Not{\alpha}_m \ ]^{\intercal},
\psp
\beta = \gal_2^{\intercal} \GLm \gal_2,
\label{DEFINITION_GAL_1}
\ee
\[
\bQ
= \left[ \ba{cc} \frac{1}{\sigma} & -\frac{\gamma}{\sigma} \\ 
0 & 1 \ea \right],
\psp 
\bF := [ \ \GSg \Bc \ \ \get \ ],
\psp
\bH = \left[ \ba{cc} \beta \sigma & 0 \\ 0 & 0 \ea \right].
\]
Partition $\bJ$ as
\[
\bJ =
\left[ \ba{cc}
\bJ_{11} 
& 
\bJ_{21}^{\intercal}
\\
\bJ_{21}
&
\bJ_{22}
\ea \right],
\]
where $\bJ_{11}$ is of size $m \times m$, while $\bJ_{22}$ is of
size $(2n+1) \times (2n+1)$. Then
\begin{align}
\bJ_{11} 
&=
\bL_1 \GLm \bL_1^{\intercal}, 
\\
\nonumber
\bJ_{21}
&=
\bF \bQ \bL_2 \GLm \bL_1^{\intercal}
\\
\nonumber
\bJ_{22}
&=
\bF
\bQ 
( \bL_2 \GLm \bL_2^{\intercal} - \bH )
\bQ^{\intercal}
\bF^{\intercal}
+ \beta \GSg,  
\end{align}
where 
\begin{align}
\bL_1 &:= \bL \bP, 
\label{DEFINITION_L1}
\\
\nonumber 
\bL_{2} &:=
\left[ 
\ba{c} 
\gal^{\intercal}_1 \\ 
\gal^{\intercal}_2 
\ea 
\right]
=
\left[ 
\ba{ccccc} 
0 & \Not{\alpha}_1 & 2 \Not{\alpha}_2 & \cdots & m \Not{\alpha}_m \\ 
\Not{\alpha}_1 & 2 \Not{\alpha}_2 & \cdots & m \Not{\alpha}_m & 0 
\ea \right] .
\end{align}
\label{THM_1}
\end{Thm}

\noindent
{\bf Proof:} See Appendix \ref{PROOF_OF_THM1}. 
\feop

\begin{Rem}
Expressions given by Theorem \ref{THM_1} allow us to compute 
$\bJ$ in a simple way. To the best of our knowledge there is no
similar expressions in the literature allowing this computational
advantage. 
\end{Rem}

The matrices 
$\bQ,\bH,\GLm,\bL_1,\bL_2$ and $\beta$ share an interesting 
property. They depend
only on the true parameter vector $\Not{\vth}$ and the 
second order statistics (consisting of $\gamma$ and $\sigma$) of the
stochastic process $w(t,\Not{\gth}) = \Bc^{\intercal} \Bx(t)$.
These quantities remain 
constant so long  $\gamma$ and $\sigma$ remain constant, even 
though the input power spectral density (and thus $\GSg$) may 
vary. This is due to the fact that the estimation accuracy of the static 
nonlinearity depends only on the amplitude distribution of $w(t)$,
regardless of $\GSg$ (or equivalently, $\Phi$).
This observation plays a key role in the sequel, and is 
formalized via the following definition.

\begin{Defn}
A quantity is called $w$-dependent if it is a function of 
$\Not{\vth}$, $\sigma$ and $\gamma$ only.
\end{Defn} 

The expressions given in Theorem \ref{THM_1} 
may not seem appealing from the point of view of setting  
up an optimization problem for input design that can be solved
in a tractable manner. The next result is more attractive in 
that regard. 

\begin{Thm}
\label{THM_2}
The determinant of $\bJ$ is given by
\be
\det(\bJ) = \frac{ \beta^{2n} r_1^2}{\sigma} \det(\bJ_{11}) 
\det(\GSg).
\label{J_r2_0}
\ee
where $r_1 = 
\gal_1^{\intercal} \gup ( \gup^{\intercal} \GLm^{-1} \gup )^{-1/2}$.
\end{Thm}

\noindent
{\bf Proof:} See Appendix \ref{PROOF_DET_INVERSE}
\feop

\begin{Rem} The expression of $\det(\bJ)$ in \cf{J_r2_0} has some
nice structure.  The factor 
\be
f :=\beta^{2n} r_1^2 \det(\bJ_{11}) / \sigma
\label{DEFINITION_F}
\ee 
is $w$-dependent, and remains constant when the statistics of
$w(t,\gth_0)$ remain invariant. On the other 
hand it is well-known from the literature on the input design for linear
systems that we can parameterize $\det(\GSg)$ in a convex manner 
using a finite number of parameters. When the mean $\eta_u$ of the 
input is kept fixed, then the above facts let us  solve the
D-optimal design problem for Wiener models via an one dimensional 
search in $\sigma$. To emphasize the $w$-dependence of $f$ we write
it as $f(\gamma,\sigma)$. When we consider a situation where 
$\gamma$ is fixed and known, then we write it as $f(\sigma)$.  
\label{REM_8}
\end{Rem}

\begin{Rem} Note that $\bJ$ is singular when $r_1 = 0$. 
This means that the normalization of the form described in Assumption
\ref{ASM2} ensures identifiability  (and thus a non-singular information
matrix) only when 
\be
0 \ne \gal_1^{\intercal} \gup = \upsilon_1 \Not{\alpha}_1 + 
2 \upsilon_2 \Not{\alpha}_2 + \cdots +
m \upsilon_m \Not{\alpha}_m, 
\label{IDENTIFIABILITY}
\ee
see the definition of $r_1$ in the statement of Theorem \ref{THM_2}.  
We can easily construct a case where \cf{IDENTIFIABILITY} fails to hold. 
That is
$\gup = (1,0,\ldots,0)$. It is straightforward to see why this 
choice leads to lack of identifiability:
 it still allows us to 
simultaneously vary the gain 
of the linear subsystem and the factors $\{ \alpha_k \}_{k=1}^m$,
while the constraint \cf{CONSTRAINT} is satisfied.

By imposing the constraint \cf{CONSTRAINT} we restrict the search space
to the hyperplane
\[
\mathcal{H} = \left \{ (\alpha_0, \alpha_1, \cdots \alpha_m) :
\sum_{k=0}^m \alpha_k \upsilon_k = 1 \right\}
\]
By assumption, $(\Not{\alpha}_0,
\Not{\alpha}_1,\cdots,\Not{\alpha}_m) \in \mathcal{H}$.
The model is identifiable when $\mathcal{H}$ intersects with the 
manifold
\[
\mathcal{M} = \{ (  \Not{\alpha}_0,
\varrho \Not{\alpha}_1,\cdots, \varrho^m \Not{\alpha}_m )
: \varrho \ne 0 \}
\]
only at the point $(\Not{\alpha}_0,
\Not{\alpha}_1,\cdots,\Not{\alpha}_m)$, which corresponds to $\varrho =1$. 
We have local identifiability at 
$(\Not{\alpha}_0,\Not{\alpha}_1,\cdots,\Not{\alpha}_m)$ only if  
$\mathcal{M}$ is not oriented along $\mathcal{H}$ at 
$(\Not{\alpha}_0,
\Not{\alpha}_1,\cdots,\Not{\alpha}_m)$, i.e.,
$\varrho =1$. In other words, we do not want the directional derivative 
$(0,2 \Not{\alpha}_1, \cdots, m \Not{\alpha}_m ) =: \gal_1$
of $\mathcal{M}$ at $\varrho = 1$ to be perpendicular to $\gup$, 
which is identical to \cf{IDENTIFIABILITY}. 
\label{INFOMATRIX_AND_IDENTIFIABILITY}
\end{Rem}
 

\section{Conclusions}

We have presented several new results on the analysis of Wiener
model identification using Gaussian input processes. One of the main 
results in this paper is Theorem \ref{THM_1}, which gives a closed form
expression of the associated information matrix $\bJ$. 
This expression holds 
under very generic assumptions on the model structure. In addition, 
unlike other similar formulae available in the literature, our 
expression for $\bJ$ is easy to compute. This aspect makes it attractive
in input design problems. 
Theorem \ref{THM_2} gives a simple expression for the determinant of 
$\bJ$. These expressions are particularly useful in experiment design,
see \cite{Mahata15} for details.


\appendix


\section{Proof of Lemma \ref{LEM2}}
\label{PROOF_OF_LEM2}

By definition of $\bP$ in \cf{DEFINITION_PERMUTATION} we have
$\bP \bP^{\intercal}  = \bI$. Using this in \cf{EXPRESSOIN_GAL}
gives 
\be
\bar{\gal} = \bP^{\intercal} \bL^{\intercal} \gal.
\label{EXPRESSOIN_BAR_GAL}
\ee
Using \cf{EXPRESSOIN_BAR_GAL} and 
the definition of $\Bz(t,\gth)$ in \cf{ExpressionJ}
in \cf{MODEL}  we have
\be
M(\vth,\Bu_t) = \bar{\gal}^{\intercal} \Bz(t,\gth) = 
\gal^{\intercal} \bL \bP \Bz(t,\gth).
\label{EQN6}
\ee
Hence
\be
\frac{ \partial  M(\Not{\vth},\Bu_t) }{ \partial \gal } = 
\bL \bP \Bz(t, \Not{\gth}).
\psp
\ee
Also using the definition of $\Bz(t,\gth)$ in \cf{ExpressionJ}
and differentiating $M(t,\vth)$ in \cf{EQN6} with respect to 
$\gth$ we get
\begin{align}
\nonumber
\frac{ \partial M(\Not{\vth}, \Bu_t) }{ \partial \gth }  
&= 
\frac{ \partial w( t,\Not{\vth} ) }{ \partial \gth } \ 
\Not{\gal}^{\intercal} \bL \bP \left[ \ba{c}
0 \\ 1 \\ 2 w(t,\Not{\gth}) \\ \vdots \\
m \{ w(t,\Not{\gth}) \}^{m-1}
\ea \right]
\\
&=
\frac{ \partial w( t,\Not{\vth} ) }{ \partial \gth }  \
\gal_2^{\intercal} \Bz(t,\Not{\gth}),
\label{A8}
\end{align}
where the last equality follows from the definition of $\gal_2$ in 
\cf{DEFINITION_GAL_2} and the definition of $\Bz(t,\gth)$ in \cf
{ExpressionJ}. The proof for the expression of $\Bv_t$ in 
\cf{Expression_vt} will be complete if we can show
\be
\frac{ \partial w( t,\Not{\vth} ) }{ \partial \gth } = 
\frac{ \partial G(z,\Not{\gth})}{\partial \gth} u(t) =
\Bx(t).
\label{EQN21}
\ee
This is done next.
By differentiating $G$ with respect to $\Ba$, $\Bg$ and $g_0$  we get
\begin{subequations}
\begin{align}
\nonumber
\frac{\partial G(z,\gth)}{\partial \Ba}
&=
- (z \bI - \bA_1)^{-1} \Bb_1 g_0  \\
&
\hspace{-5mm}
- (z \bI - \bA_1)^{-1} \Bb_1 (\Bg - \Ba g_0)^{\intercal}(z \bI - \bA_1)^{-1} \Bb_1,
\\
\frac{\partial G(z,\gth)}{\partial \Bg}
&= (z \bI - \bA_1)^{-1} \Bb_1,  \\
\frac{\partial G(z,\gth)}{\partial g_0}
&=
1   - \Ba^{\intercal} (z \bI - \bA_1)^{-1} \Bb_1. 
\end{align}
\label{G1}
\end{subequations}
Using \cf{G1} and \cf{A2b2C2} it can be verified by direct calculations  that 
\be
\frac{\partial G(z,\Not{\gth})}{\partial \gth}
= 
(\bI - \bA z^{-1} )^{-1} \Bb,
\label{AnotherG1}
\ee
implying \cf{EQN21}.

To show $w(t,\Not{\gth}) = \Bc^{\intercal} \Bx(t)$ verify 
from \cf{A5a} and \cf{G1} that 
\[
G(z,\Not{\gth}) = 
[ \ {\bf 0}^{\intercal} \ \ \Not{\Bg}^{\intercal}
\ \ \Not{g}_0 \ ]
\frac{\partial G(z,\Not{\gth})}{\partial \gth} 
= \Bc^{\intercal} (\bI - \bA z^{-1} )^{-1} \Bb. 
\]



\section{Proof of Theorem \ref{THM_1}}
\label{PROOF_OF_THM1}

Since $\GSg$ is positive definite, $\GSg \Bc \ne 0$. Hence
there exists a full column rank 
$(2n+1)\times(2n)$ 
matrix
$\bC$ such that the column 
space of $\bC$ is the orthogonal complement of $\GSg \Bc$,
i.e., $\bC^{\intercal} \GSg \Bc = 0$. Hence
\be
\left[ \ba{c} \Bc^{\intercal} \\ \bC^{\intercal} \ea \right]
\
\GSg 
\
[ \ \Bc \ \ \bC \ ]
=
\left[ \ba{cc}
\sigma & 0 \\ 0 & \GSg_1
\ea \right],
\label{A21}
\ee
The block diagonal structure of the matrix in the right hand side 
of \cf{A21}
ensures that by premultiplying $\Bx$ by 
$[ \ \Bc \ \ \bC \ ]^{\intercal}$ we 
get two mutually uncorrelated components 
$\Bc^{\intercal} \Bx$ and 
\[
\Bx_1 := \bC^{\intercal} \Bx,
\] 
with 
\begin{align}
\nonumber
\ggm &:= \Ex \{ \Bx_1 \} = \bC^{\intercal} \get, 
\\
\GSg_1 &:= 
\Ex \{ [\Bx_1 - \ggm] [\Bx_1 - \ggm]^{\intercal} \} = 
\bC^{\intercal} \GSg \bC.
\end{align}
Because $\Bx$ is a Gaussian random vector, we conclude that 
$[ \ \Bc^{\intercal} \Bx \ \ \Bx_1^{\intercal} \ ]^{\intercal}$ too
is a jointly Gaussian random vector. Since uncorrelated Gaussian
variables are independent, $\Bc^{\intercal} \Bx$ and $\Bx_1$ 
are  mutually independent.

Define the $(m+2n+1) \times (m+2n+1)$ matrix
\be
\bT = 
\left[
\ba{cc} 
\bI & 0 \\
0 & \Bc^{\intercal} \\
0 & \bC^{\intercal}
\ea \right],
\label{defT}
\ee
where the identity matrix appearing
in \cf{defT} in the north-west corner is of size $m \times m$.
Premultiplying $\Bv_t$ in \cf{ExpressionJ} by $\bT$
we note that
\be
\bT
\Bv_t
=
\left[ \ba{c} 
\bL_1 \Bz(t, \Not{\gth}) \\
\Bc^{\intercal} \Bx(t) \gal_2^{\intercal} \Bz(t,\Not{\gth}) \\
\bC^{\intercal} \Bx(t) \gal_2^{\intercal} \Bz(t,\Not{\gth}) \ea 
\right].
\ee
From Lemma \ref{LEM2} recall that  
$\Bc^{\intercal} \Bx(t) = w(t,\Not{\gth})$. Then from 
the definition of $\Bz(t,\gth)$ in \cf{ExpressionJ}, 
the definitions $\gal_1$ and $\gal_2$ in \cf{DEFINITION_GAL_1} and
\cf{DEFINITION_GAL_2} we have 
\[
\Bc^{\intercal} \Bx(t) \gal_2^{\intercal} \Bz(t,\Not{\gth})
= 
w(t,\Not{\gth}) 
\gal_1^{\intercal} \Bz(t,\Not{\gth}).
\]
In addition, $\bC^{\intercal} \Bx(t) = \Bx_1(t)$. Hence
\be
\bT \Bv_t
=
\left[ \ba{c} 
\bL_1 \Bz(t, \Not{\gth}) \\
\gal_1^{\intercal} \Bz(t,\Not{\gth}) \\
\Bx_1(t) \gal_2^{\intercal} \Bz(t,\Not{\gth}) \ea 
\right].
\label{eqLem3}
\ee
Since $\Bx_1(t)$ is independent of 
$w(t,\Not{\gth}) = \Bc^{\intercal} \Bx(t)$, it
is also independent of $\Bz(t,\Not{\gth})$, see 
\cf{ExpressionJ}. Using this
and \cf{eqLem3} we get 
\begin{align}
\nonumber
& \bT \bJ \bT^{\intercal}
=
\Ex \left \{ 
\left[ \bT \Bv_t \right]
\left[ \bT \Bv_t \right]^{\intercal}
\right \}
\\
&=
\left[ \ba{ccc}
\bL_1 \GLm \bL_1^{\intercal} & \bL_1 \GLm \gal_1 & 
\bL_1 \GLm \gal_2 \ggm^{\intercal} \\
\gal_1^{\intercal} \GLm \bL_1^{\intercal} & \gal_1^{\intercal} \GLm \gal_1 & \gal_1^{\intercal} \GLm \gal_2 \ggm^{\intercal} \\
\ggm \gal_2^{\intercal} \GLm \bL_1^{\intercal} & 
\ggm \gal_2^{\intercal} \GLm \gal_1 & 
 \gal_2^{\intercal} \GLm \gal_2 ( \ggm \ggm^{\intercal} + \GSg_1)
\ea \right].
\label{A31}
\end{align}

Define the vector $\Bd$ and the $(2n+1)\times(2n)$ matrix $\bD$ 
by partitioning the 
inverse
\be
\left[ 
\ba{c} 
\Bc^{\intercal} \\ \bC^{\intercal} 
\ea 
\right]^{-1}
=
[ \ \Bd \ \ \bD \ ].
\label{A33}
\ee
Then \cf{defT} and \cf{A31} imply
\begin{align}
\nonumber
\bJ
&= 
\left[ \ba{ccc}
\bI & 0 & 0 \\
0 & \Bd & \bD 
\ea \right]
(\bT \bJ \bT^{\intercal}) 
\left[ \ba{cc}
\bI & 0 \\
0 & \Bd^{\intercal} \\
0 & \bD^{\intercal}  
\ea \right].
\end{align}
Using expression of $\bT \bJ \bT^{\intercal}$ in \cf{A31} we
get
\begin{subequations}
\begin{align}
\bJ_{11} 
&=
\bL_1 \GLm \bL_1^{\intercal}, 
\\
\bJ_{21} 
&=
[ \ \Bd \ \ \bD \ggm \ ] \ \bL_2 \GLm \bL_1^{\intercal} 
\\
\bJ_{22} 
&=
[ \ \Bd \ \ \bD \ggm \ ] \ \bL_2 \GLm \bL_2^{\intercal} \
[ \ \Bd \ \ \bD \ggm \ ]^{\intercal} 
+ 
\beta
\bD \GSg_1 \bD^{\intercal}.
\label{A35}
\end{align}
\label{eqThm1}
\end{subequations}
Now from \cf{A21} and \cf{A33} we obtain
\be
\GSg = [ \ \Bd \ \ \bD \ ] 
\left[ \ba{cc} \sigma & 0 \\ 0 & \GSg_1 \ea \right]
\left[ \ba{c}
\Bd^{\intercal} \\ \bD^{\intercal} 
\ea \right]
= \Bd \sigma \Bd^{\intercal} + \bD \GSg_1 \bD^{\intercal}. 
\label{A36}
\ee
By definition of $\Bd$ and $\bD$ in \cf{A33} we know
\[
\left[
\ba{c}
\Bc^{\intercal}
\\
\bC^{\intercal}
\ea
\right]
[ \ \Bd \ \ \bD \ ]
=
\left[ \ba{cc}
1 & 0 \\ 0 & \bI 
\ea \right],
\]
and this implies
$
\bC^{\intercal} \Bd = 0, \psp \Rightarrow
\Bd = k \GSg \Bc.
$
In addition,
$
1= \Bc^{\intercal} \Bd = 
k \Bc^{\intercal} \GSg \Bc = 
k \sigma
$.
Consequently,
\be
\Bd = \frac{1}{\sigma} \GSg \Bc.
\label{B40}
\ee
On the other hand
\be
\bI
=
[ \ \Bd \ \ \bD \ ]
\left[
\ba{c}
\Bc^{\intercal}
\\
\bC^{\intercal}
\ea
\right]
=
\Bd \Bc^{\intercal} + \bD \bC^{\intercal}
=
\frac{1}{\sigma} \GSg \Bc \Bc^{\intercal} + \bD \bC^{\intercal},
\label{B41a}
\ee
Now multiply both sides of \cf{B41a} by $\get$ to get
\be
\get - \frac{\gamma}{\sigma} \GSg \Bc = \bD \ggm
\label{B41b}
\ee
From \cf{B40} and \cf{B41b} it follows that
\[ 
[ \ \Bd \ \ \bD \ggm \ ] 
=
\bF \bQ.
\ \ 
\]
Now we use \cf{A36}, \cf{B40}, and \cf{B41b} in \cf{eqThm1} to 
eliminate 
$\Bd$ and $\bD$ from the expressions of $\bJ_{12}$ and $\bJ_{22}$. We have
\begin{align}
\nonumber
\bJ_{21}
&=
\bF \bQ \bL_2 \GLm \bL_1^{\intercal}
\\
\nonumber
\bJ_{22}
&=
\bF \bQ \bL_2 \GLm \bL_2^{\intercal} \bQ^{\intercal}
\bF^{\intercal}
+ \beta( \GSg - \GSg \Bc \Bc^{\intercal} \GSg/\sigma) 
\\
\nonumber
&=
\beta \GSg +
\bF \bQ \bL_2 \GLm \bL_2^{\intercal} \bQ^{\intercal} \bF^{\intercal}
- 
\bF \bQ \bH \bQ^{\intercal} \bF^{\intercal}
\\
\nonumber
&=
\bF \bQ 
( \bL_2 \GLm \bL_2^{\intercal} - \bH )
\bQ^{\intercal}
\bF^{\intercal}
+ \beta \GSg.  
\end{align}


\section{Proof of Theorem \ref{THM_2}}
\label{PROOF_DET_INVERSE}

\subsection{Some Schur complement expressions}

\begin{Lem}
The Schur complement $\bJ_{22} - \bJ_{21} 
\bJ_{11}^{-1} \bJ_{21}^{\intercal}$ admits an expression
\begin{align}
\nonumber
& \bJ_{22} - \bJ_{21} \bJ_{11}^{-1} \bJ_{21}^{\intercal}
\\
\nonumber
&=
\beta \GSg + 
\bF \bQ [
\bL_2 \gup (\gup^{\intercal} \GLm^{-1} \gup)^{-1} \gup^{\intercal}
\bL_2^{\intercal}  - \bH ] 
\bQ^{\intercal} \bF^{\intercal}.
\end{align}
\label{LEM_4}
\end{Lem}

\noindent
{\bf Proof:}
In this proof we let $\GGm$ be the Cholesky factor of $\GLm$,
i.e., $\GLm = \GGm \GGm^{\intercal}$.
From the expressions of $\bJ_{11}$,
$\bJ_{21}$ and $\bJ_{22}$ in Theorem \ref{THM_1} it follows that 
\be
\bJ_{22} - \bJ_{21} \bJ_{11}^{-1} \bJ_{21}^{\intercal}
=
\beta \GSg + 
\bF \bQ [
\bL_2 \GPi
\bL_2^{\intercal}  - \bH ] 
\bQ^{\intercal} \bF^{\intercal},
\label{eq:LEM_4}
\ee
where we define
\begin{align}
\nonumber
\GPi 
&= 
\GLm - \GLm \bL_1^{\intercal} ( \bL_1 \GLm \bL_1^{\intercal} )^{-1}
\bL_1 \GLm
\\
&= 
\GGm [ \bI - \GGm^{\intercal} \bL_1^{\intercal} 
( \bL_1 \GGm \GGm^{\intercal} \bL_1^{\intercal} )^{-1}
\bL_1 \GGm ] \GGm^{\intercal}.
\label{GPI}
\end{align}
However, the matrix 
$\bar{\GPi} := \bI - \GGm^{\intercal} \bL_1^{\intercal} 
( \bL_1 \GGm \GGm^{\intercal} \bL_1^{\intercal} )^{-1}
\bL_1 \GGm$ is the orthogonal projection operator onto 
the nullspace of $\bL_1 \GGm$. 

From \cf{DEFINITION_GUP}, \cf{DEFINITION_PERMUTATION} and the 
definition of $\bL_1$ in \cf{DEFINITION_L1} verify that
that $\bL_1 \gup = \bL \bP \gup = 0$. This means
\[
\bL_1 \GGm \GGm^{-1} \gup = 0,
\]
i.e. the vector $\GGm^{-1} \gup$ spans the one dimensional nullspace
of $\bL_1 \GGm$. Hence $\bar{\GPi}$ is also 
the orthogonal projection operator onto 
the span of $\GGm^{-1} \gup$. Hence
\[
\bar{\GPi}
=
\GGm^{-1} \gup ( \gup^{\intercal} \GLm^{-1} \gup )^{-1} 
\gup^{\intercal} \GGm^{- \intercal}.
\]
Substituting this expression in \cf{GPI} gives
\[
\GPi = \gup ( \gup^{\intercal} \GLm^{-1} \gup )^{-1} 
\gup^{\intercal},
\]
which upon substitution in \cf{eq:LEM_4} yields the desired result.
\feop

Define
\be
r_i := \gal_i^{\intercal} \gup
( \gup^{\intercal} \GLm^{-1} \gup )^{-1/2}, \psp i=1,2.
\label{def_r_i}
\ee
Note that
\be
\bL_2 \gup (\gup^{\intercal} \GLm^{-1} \gup)^{-1} \gup^{\intercal}
\bL_2^{\intercal} 
=
\left[ \ba{c} r_1 \\ r_2 \ea \right]
\left[ \ba{c} r_1 \\ r_2 \ea \right]^{\intercal},
\label{L2_GUP_}
\ee
see the definition of $\bL_2$ in Theorem \ref{THM_1}.
When $r_2 = 0$  the matrix
$\bL_2 \gup (\gup^{\intercal} \GLm^{-1} \gup)^{-1} \gup^{\intercal}
\bL_2^{\intercal}  - \bH$ is of rank $1$. Then the calculations
turn out to be quite different from the case where $r_2 \ne 0$.

\begin{Lem}
If $r_2 = 0$ then 
\begin{align}
\det(\bJ_{22} - \bJ_{21} \bJ_{11}^{-1} \bJ_{21}^{\intercal})
&=
\frac{r_1^2}{\beta \sigma} \det(\beta \GSg), 
\label{DET_LEM_5}
\end{align}
\label{LEM_5}
\end{Lem}

\noindent
{\bf Proof:}
When $r_2 = 0$ then using \cf{def_r_i},
definition of $\bQ$ in Theorem \ref{THM_1} and
the expressions  given by Lemma 
\ref{LEM_4} we get
\begin{align}
\nonumber
& \bJ_{22} - \bJ_{21} \bJ_{11}^{-1} \bJ_{21}^{\intercal}
\\
\nonumber
&=
\beta \GSg + 
\bF \bQ 
\left [
\ba{cc} r_1^2 - \beta \sigma & 0 \\ 0 & 0 \ea \right]
\bQ^{\intercal} \bF^{\intercal}
\\
\nonumber
&=
\beta \GSg + 
\bF 
\left[ \ba{cc} \frac{1}{\sigma} & -\frac{\gamma}{\sigma} \\ 
0 & 1 \ea \right] 
\left [
\ba{cc} r_1^2 - \beta \sigma & 0 \\ 0 & 0 \ea \right]
\bQ^{\intercal} \bF^{\intercal}
\\
\nonumber
&=
\beta \GSg + 
\bF 
\left [
\ba{cc} r_1^2/\sigma - \beta & 0 \\ 0 & 0 \ea \right]
\left[ \ba{cc} 
\frac{1}{\sigma} & 0 \\ 
-\frac{\gamma}{\sigma} & 1
\ea \right] 
\bF^{\intercal}
\\
\nonumber
&=
\beta \GSg + 
\bF 
\left [
\ba{cc} r_1^2/\sigma^2 - \beta/\sigma & 0 \\ 0 & 0 \ea \right]
\bF^{\intercal}
\\
&=
\beta \GSg + 
(r_1^2/\sigma^2 - \beta/\sigma) \GSg \Bc \Bc^{\intercal} \GSg.
\label{SCHUR_R_0}
\end{align}
In this proof we write 
\[
q = r_1^2/\sigma^2 - \beta/\sigma
\] 
compactly.  From \cf{SCHUR_R_0} we have
\begin{align}
\nonumber
& \det( \bJ_{22} - \bJ_{21} \bJ_{11}^{-1} \bJ_{21}^{\intercal} )
=
\det \left( 
\beta \GSg + 
q \GSg \Bc \Bc^{\intercal} \GSg
\right)
\\
\nonumber
&=
\det ( \beta \GSg ) 
\det \left( \bI + 
\frac{q}{\beta}  \Bc \Bc^{\intercal} \GSg
\right)
\\
\nonumber
&=
\det ( \beta \GSg ) 
\det \left( 1 + 
\frac{q}{\beta}  \Bc^{\intercal} \GSg \Bc
\right)
\\
\nonumber
&=
\det ( \beta \GSg ) 
\det \left( 1 + \frac{q \sigma}{\beta} \right)
\end{align}
Substituting the value of $q$ we get \cf{DET_LEM_5}.

\feop

\begin{Lem}
Suppose that $r_2 \ne 0$. Then 
\begin{align}
\det(\bJ_{22} - \bJ_{21} \bJ_{11}^{-1} \bJ_{21}^{\intercal})
&=
\frac{r_1^2}{\beta \sigma} \det(\beta \GSg), 
\label{DET_LEM_6}
\\
\nonumber
(\bJ_{22} - \bJ_{21} \bJ_{11}^{-1} \bJ_{21}^{\intercal})^{-1}
&=
\left[
\frac{1}{r_1^2} - \frac{1}{\beta \sigma} 
\left( \frac{r_2 \gamma}{r_1} - 1 \right)^2
\right]
\Bc \Bc^{\intercal} \\
& \hspace{-2cm} +
\left( 
\frac{r_2}{r_1} \Bc \get^{\intercal} -  \bI 
\right) 
[\beta \GSg]^{-1} 
\left( 
\frac{r_2}{r_1} \Bc \get^{\intercal} -  \bI 
\right)^{\intercal}.
\label{INV_LEM_6}
\end{align}
\label{LEM_6}
\end{Lem}

\noindent
{\bf Proof:}  Define 
\be
\bB := 
\left[
\bQ
\left(
\frac
{\bL_2 \gup \gup^{\intercal} \bL_2^{\intercal}}  
{(\gup^{\intercal} \GLm^{-1} \gup)^{-1}} 
-
\bH
\right) 
\bQ^{\intercal}
\right]^{-1}
\label{def_B}
\ee
Recall that $\zeta = r_1/r_2$. Hence from \cf{L2_GUP_} we get
\begin{align}
\nonumber
& \hspace{-5mm}
\left(
\frac
{\bL_2 \gup \gup^{\intercal} \bL_2^{\intercal}}  
{(\gup^{\intercal} \GLm^{-1} \gup)^{-1}} 
-
\bH
\right)^{-1} 
\\
\nonumber
&=
-
\frac{1}{ \beta \sigma  }
\left[ \ba{cc} 
1  & -r_1 / r_2 \\ -r_1 / r_2 & r_1^2/r_2^2 - \beta \sigma /r_2^2 
\ea \right]
\\
\nonumber
&=
-\frac{1}{ \beta \sigma  }
\left[ \ba{cc} 
1  & -\zeta \\ -\zeta & \zeta^2 - \beta \sigma /r_2^2 
\ea \right].
\end{align}
Hence by definition of $\bQ$, see Theorem \ref{THM_1}, we get
\bea
\nonumber
\beta \bB 
&=& 
-\frac{1}{  \sigma  }
\left[ \ba{cc} \sigma & 0 \\ \gamma & 1 \ea \right]
\left[ \ba{cc} 
1  & -\zeta \\ -\zeta & \zeta^2 - \beta \sigma /r_2^2 
\ea \right]
\bQ^{-1}
\\
\nonumber
&=& 
-\frac{1}{ \beta \sigma  }
\left[ \ba{cc} 
\sigma  & - \sigma \zeta \\ \gamma-\zeta & -\zeta( \gamma-\zeta)  - \beta \sigma / r_2^2 
\ea \right]
\left[ \ba{cc} \sigma & \gamma \\ 0 & 1 \ea \right]
\\
\nonumber
&=& 
-\frac{1}{  \sigma  }
\left[ \ba{cc} 
\sigma^2  & \sigma (\gamma - \zeta) \\ 
\sigma(\gamma-\zeta) & ( \gamma-\zeta)^2  - \beta \sigma / r_2^2 
\ea \right]
\\
&=& 
-
\left[ \ba{cc} 
\sigma  & \gamma - \zeta \\ 
\gamma-\zeta & \frac{ (\gamma-\zeta)^2 }{\sigma}  - \beta / r_2^2 
\ea \right].
\label{betaB}
\eea
Taking determinant we have
\be
\det(\beta \bB) = - \frac{
\beta \sigma}{r_2^2}.
\label{det_beta_B}
\ee
On the other hand, recall that 
$\bF = [ \ \GSg \Bc \ \ \get \ ]$. Hence using 
\cf{gamma_sigma} we get
\be
\bF^{\intercal} \GSg^{-1} \bF
=
\left[ \ba{cc}
\sigma & \gamma \\ \gamma & \get^{\intercal} \GSg^{-1} \get 
\ea \right].
\label{Ft_SigInv_F}
\ee
Combining \cf{betaB} and \cf{Ft_SigInv_F} we get
\be
\beta \bB + \bF^{\intercal} \GSg^{-1} \bF
= 
\left[
\ba{cc}
0 & \zeta \\ \zeta & \get^{\intercal} \GSg^{-1} \get + \frac{\beta}{r_2^2} - 
\frac{(\gamma - \zeta)^2}{\sigma} \ea \right]
\label{betaB_plus_Ft_SigInv_F}
\ee
Taking determinant we have
\be
\det( \beta \bB + \bF^{\intercal} \GSg^{-1} \bF )
=
-\zeta^2
\label{det_betaB_plus_Ft_SigInv_F}
\ee
Now using Lemma \ref{LEM_4} and \cf{def_B} we know
\be
\bJ_{22} - \bJ_{21} \bJ_{11}^{-1} \bJ_{21}^{\intercal}
= 
\beta \GSg + \bF \bB^{-1} \bF^{\intercal}
\label{Schur_Formula}
\ee
Hence 
\begin{align}
\nonumber
& \det( \bJ_{22} - \bJ_{21} \bJ_{11}^{-1} \bJ_{21}^{\intercal} )
\\
\nonumber
&=
\det ( \beta \GSg + 
\bF \bB^{-1} \bF^{\intercal}
)
\\
\nonumber
&=
\det(\beta \GSg)
\det ( \bI + 
\GSg^{-1} \bF
(\beta \bB)^{-1} 
\bF^{\intercal}
)
\\
\nonumber
&=
\det(\beta \GSg)
\det ( \bI + 
\bF^{\intercal} \GSg^{-1} \bF
\{ \beta \bB \}^{-1}
)
\\
\nonumber
&=
\frac{\det(\beta \GSg)}{\det( \beta \bB)}
\det \left( 
\beta \bB  + 
\bF^{\intercal} \GSg^{-1} \bF
\right )
\\
\nonumber
&=
\det(\beta \GSg) 
\frac{r_2^2 \zeta^2}{\beta \sigma}
=
\det(\beta \GSg) 
\frac{r_1^2}{\beta \sigma}.
\end{align}

\subsection{Proof of the formula for $\det(\bJ)$}

Using Schur's determinant formula we know
\begin{align}
\det(\bJ) = \det(\bJ_{11}) 
\det( \bJ_{22} - \bJ_{21} \bJ_{11}^{-1} \bJ_{21}^{\intercal} ).
\label{eq:THM_2}
\end{align}
The result of Theorem \ref{THM_2} is immediate from \cf{eq:THM_2}
once we use the expression for 
$\det( \bJ_{22} - \bJ_{21} \bJ_{11}^{-1} \bJ_{21}^{\intercal} )$
given by \cf{DET_LEM_6}.

\bibliographystyle{plain}  
\bibliography{sam}         

\begin{thebibliography}{10}

\bibitem{IR-EE-RT_2007:003}
M.~Barenthin, X.~Bombois, H.~Hjalmarsson, and G.~Scorletti.
\newblock Identification for control of multivariable systems: Controller
  validation and experiment design via {LMI}s.
\newblock {\em Automatica}, 44:3070--3078, Dec 2008.

\bibitem{IR-EE-RT_2006:026}
M.~Barenthin and H.~Hjalmarsson.
\newblock Identification and control: Joint input design and
  $\mathcal{H}_{\infty}$ state feedback with ellipsoidal parametric uncertainty
  via {LMI}s.
\newblock {\em Automatica}, 44(2):543--551, Feb 2008.

\bibitem{Bombois06}
X.~Bombois, G.~Scorletti, M.~Gevers, P.~M. J.~Van den {H}of, and R.~Hildebrand.
\newblock Least costly identification experiment for control.
\newblock {\em Automatica}, 42:3:1651--1662, 2006.

\bibitem{DeCock_13}
A.~De~Cock, M.~Gevers, and J.~Schoukens.
\newblock A preliminary study on optimal input design for nonlinear systems.
\newblock In {\em Decision and Control (CDC), 2013 IEEE 52nd Annual Conference
  on}, pages 4931--4936, Dec 2013.

\bibitem{DeCock_14}
A.~De~Cock, M.~Gevers, and J.~Schoukens.
\newblock D-optimal input design for {FIR}-type nonlinear systems: A dispersion
  based approach.
\newblock 2014.
\newblock Submitted for publication.

\bibitem{Forgione_14}
M.~Forgione, X.~Bombois, P.~M.~J. Van~den Hof, and H.~Hjalmarsson.
\newblock Experiment design for parameter estimation in nonlinear systems based
  on multilevel excitation.
\newblock 2014.
\newblock European Control Conference.

\bibitem{Gevers_12}
M.~Gevers, M.~Caenepeel, and J.~Schoukens.
\newblock Experiment design for the identification of a simple {W}iener system.
\newblock In {\em Decision and Control (CDC), 2012 IEEE 51st Annual Conference
  on}, pages 7333--7338, Dec 2012.

\bibitem{Hildebrand03}
R.~Hildebrand and M.~Gevers.
\newblock Identification for control: optimal input design with respect to a
  worst-case $\nu$-gap cost function.
\newblock {\em {SIAM} journal of control and optimization}, 41:5:1586--1608,
  2003.

\bibitem{IR-EE-RT_2006:053}
H.~Hjalmarsson and H.~Jansson.
\newblock Closed loop experiment design for linear time invariant dynamical
  systems via {LMI}s.
\newblock {\em Automatica}, 44(3):623--636, Mar 2008.

\bibitem{Hjalmersson_07}
H.~Hjalmarsson and J.~M{\aa}rtensson.
\newblock Optimal input design for identification of non-linear systems:
  Learning from the linear case.
\newblock In {\em American Control Conference, 2007. ACC '07}, pages
  1572--1576, July 2007.

\bibitem{1924}
H.~Jansson and H.~Hjalmarsson.
\newblock Input design via {LMI}s admitting frequency-wise model specifications
  in confidence regions.
\newblock {\em {IEEE} Transactions on Automatic Control}, 50(10):1534--1549,
  Oct 2005.

\bibitem{Kan_08}
R.~Kan.
\newblock From moments of sum to moments of product.
\newblock {\em Journal of multivariate analysis}, 99:542--554, 2008.

\bibitem{Larsson_10}
C.~A. Larsson, H.~Hjalmarsson, and C.~R. Rojas.
\newblock On optimal input design for nonlinear {FIR}-type systems.
\newblock In {\em Decision and Control (CDC), 2010 49th IEEE Conference on},
  pages 7220--7225, Dec 2010.

\bibitem{Mahata15}
K.~Mahata, J.~Schoukens, and A.~De Cock.
\newblock Information matrix and {D}-optimal design with gaussian inputs for
  wiener model identification.
\newblock {\em Automatica}, 2015.
\newblock In press.

\bibitem{Papoulis_91}
A.~Papoulis.
\newblock {\em Probability, Random Variables, and Stochastic Processes}.
\newblock McGraw-Hill, New York, 1991.

\bibitem{Valenzuela_13}
P.~E. Valenzuela, C.~R. Rojas, and H.~Hjalmarsson.
\newblock Optimal input design for non-linear dynamic systems: A graph theory
  approach.
\newblock In {\em Decision and Control (CDC), 2013 IEEE 52nd Annual Conference
  on}, pages 5740--5745, Dec 2013.

\end{thebibliography}
\end{document}